\documentclass[aps,prl,twocolumn,floatfix,showpacs]{revtex4-1}

\usepackage{graphicx}
\usepackage{amsmath,amssymb}
\usepackage{dcolumn} 
\usepackage{bm} 
\usepackage{xcolor}
\usepackage[utf8]{inputenc}
\usepackage{soul}
\usepackage{comment}
\usepackage{mathtools}

\DeclarePairedDelimiterX\braket[2]{\langle}{\rangle}{#1 \delimsize\vert #2}

\usepackage{hyperref}
\hypersetup{%
  colorlinks=true,
  citecolor=blue,
  urlcolor=blue,
  linkcolor=blue,
  }

\begin{document}

\preprint{APS/123-QED}

\title{Electrostatic control of the trion fine structure
in transition metal dichalcogenide monolayers}

\author{Yaroslav V. Zhumagulov$^{1,2}$}
\author{Alexei Vagov$^{2}$}
\author{Dmitry R. Gulevich$^{2}$}
\author{Vasili Perebeinos$^{3}$}

\affiliation{$^{1}$University of Regensburg, Regensburg, 93040, Germany}
\affiliation{$^{2}$ITMO University, St. Petersburg 197101, Russia}
\affiliation{$^{3}$Department of Electrical Engineering, University at Buffalo, The State University of New York, Buffalo, New York 14260, USA}

\date{\today}

\begin{abstract}
Charged excitons (trions) are essential for the optical spectra in low dimensional doped monolayers (ML) of transitional metal dichalcogenides (TMDC). Using a direct diagonalization of the three-body Hamiltonian, we explore the low-lying trion states in four types of TMDC MLs. We show that the trion’s fine structure results from the interplay between the spin-valley fine structure of the single-particle bands and the exchange interaction between the composing particles. We demonstrate that by variations of the doping and dielectric environment, trion energy fine structure can be tuned, leading to anti-crossing of the bright and dark states with substantial implications for the optical spectra of TMDC ML’s.
\end{abstract}

\maketitle

The 2D geometry of TMDC ML significantly enhances the Coulomb interaction,  giving rise to a much larger exciton binding energy \cite{Mak2010, Splendiani2010, Komsa2012, Feng2012, Qiu2013, BenAmara2016} compared to  bulk semiconductors~\cite{Yu2010,Muth1997,Dvorak2013}. It also facilitates other many-body states including three-particle trions~\cite{Mak2012, Ross2014, Lui2014, Rezk2016, Zhang2014, Mouri2013, Singh2016, Scheuschner2014, Soklaski2014, Zhang2015_2} and four-particle biexcitons~\cite{Torche2021,You2015,Ye2018_,Sie2015,Plechinger2015,Hao2017}.  Both the experiment and theoretical calculations demonstrated that the lowest energy optical excitations in a doped TMDC ML are often associated with trions~\cite{Mak2012, Ross2014, Lui2014, Rezk2016, Zhang2014, Mouri2013, Singh2016, Scheuschner2014, Soklaski2014, Zhang2015_2} . A sophisticated band structure with two direct band gaps at the $K$ and  $K'=-K$ points of the Brillouin zone and the presence of the spin-valley locking effect ~\cite{Tao2019,Gmitra2017,Ciccarino2018,Wang2020}  makes it possible to realize various trionic states in TMDC ML's ~\cite{Wang2018,Torche2019,Plechinger2016,Zhumagulov2020_,Courtade2017,Hichri2020,grzeszczyk2020carrier,Arora2019,chang2020variationally,Drppel2017,Efimkin2017,Jadczak2021,Zinkiewicz2021}. Commonly, they are classified by  spin and  valley quantum numbers and can be dark and bright depending on their combination. Trionic fine structure determines the optical absorption edge and the PL spectra. In particular, the PL spectra  strongly depend on whether the ground trion state is bright or dark.

There is a controversy concerning the structure and the spectral structure of the trion states in TMDC ML's. In particular, in MoS2 ML both the theory and experiment predict various trion states at the bottom of the energy spectra. These states have different internal structures, some of them being dark and others bright. They are densely packed energetically and therefore not so easy to distinguish.  Conclusions of the theoretical analysis are also ambiguous. The calculations done under the different conditions demonstrate different results. Some works assign the ground state to a dark trion state, and others conclude that it is bright. The structure of the low-lying states is crucial since it defined the spectral properties of an ML. For example, the temperature behaviour of the PL spectrum  is strongly dependent on whether the ground state is bright or dark~\cite{Zhumagulov2020_,Golovynskyi2021,Hamby2003,Amori2018}.

Here,  we investigate the lowest-energy trions in TMDC ML’s and their dependence on material-specific parameters by obtaining an exact solution for the three-particle Hamiltonian. Our calculations demonstrate that trion energy states are controlled by the interplay of the spin-orbit splitting and many-body effects. The latter are strongly influenced by doping and the dielectric environment. The most intriguing situation arises when the spin-orbit splitting is small, and the trion states are close in energy, which leads to an anti-crossing pattern. This scenario happens in MoS2 ML's and opens a tantalizing possibility for a controllable manipulation of the optical spectra in such materials simply by varying external gate voltage in a field-effect transistor setup.

{\it  Calculations of trion states} are performed by a direct diagonalization of the three-body Hamiltonian obtained by spanning the many-body model onto the states with two electrons and a hole (we consider negatively charged trions)  $ \left| c_1 c_2 v \right\rangle =  a_{c_1}^\dagger  a_{c_2}^\dagger a_{v}^\dagger \left| 0 \right\rangle$, where $c_{1,2}$ and $v$ are single-particle electron and hole states. The Hamiltonian then reads as
\begin{align}
\label{eq:three_particle}
 {  H}_{\ \ }  =& {  H}_{0} + {  H}_{cc}  + {  H}_{cv},  \,
 {  H}_{0}= (\varepsilon_{c_1} + \varepsilon_{c_2} -\varepsilon_{v}) \delta_{c_1}^{c_1^\prime} \delta_{c_2}^{c_2^\prime}  \delta_{v}^{ v^\prime}, \notag \\
 {  H}_{cc} =&+ (W_{c_1c_2}^{c_1'c_2'}-W_{c_1c_2}^{c_2'c_1'})\delta_{v}^{v'},\notag \\
 {  H}_{cv}  =& -(W_{v'c_1}^{vc_1'}-V_{v'c_1}^{c_1'v})\delta_{c_2}^{c_2'}  -(W_{v'c_2}^{vc_2'}-V_{v'c_2}^{c_2'v})\delta_{c_1}^{c_1'} \notag \\ & +(W_{v'c_1}^{vc_2'}-V_{v'c_1}^{c_2'v})\delta_{c_2}^{c_1'}  +(W_{v'c_2}^{vc_1'}-V_{v'c_2}^{c_1'v})\delta_{c_1}^{c_2'},
\end{align}
where $\varepsilon_{c,v}$ are single-particle energies, $W$ and $V$ are the screened and bare Coulomb potential. The latter in given as $ V^{ab}_{cd} = V(\textbf{k}_a -\textbf{k}_c) \langle u_c | u_a \rangle \langle u_d | u_b \rangle$ with $\langle u_c| u_a \rangle$ the overlap of the single-particle Bloch states and $V (q) = 2\pi e^2/ q$. For the screened potential, we substitute $V(q)$ in this expression by the  Rytova-Keldysh potential \cite{rytova1967the8248,keldysh,Cudazzo2011}
\begin{align}
\label{eq:Coulomb_screened}
   W(q)=
   V(q) \begin{cases}
     \varepsilon_{env} ^{-1} (1+r_0 q)^{-1} & q-\text{intravalley} \\
   \varepsilon_{bulk}^{-1} &  q-\text{interavalley} \\
    \end{cases},
\end{align}
where "intravalley" stands for the transitions within the same valley and "intervalley" is for states in different valleys. For encapsulating material, the effective dielectric constant $\varepsilon_{env} = (\varepsilon_{2}+\varepsilon_{1})/2$ is the average of dielectric constants on both sides of the ML, and the screening length is  $r_0=\varepsilon_{bulk} d/2$ with $d$ being the ML width \cite{Berkelbach2013,Cho2018}. For the single-particle states, we assume the massive $k\cdot p$ Dirac model with the Hamiltonian ~\cite{Xiao2012}
\begin{align}
    H_D&=  \hbar v_{F} \big( \tau k_x\sigma_x+k_y\sigma_y \big) + \frac{\Delta}{2} s_0\otimes\sigma_z \notag \\
    &+\frac{1}{2} \tau s_{z} \otimes \big( [\lambda_c - \lambda_v] \sigma_{z} + [\lambda_c - \lambda_v] \sigma_0 \big),
\label{HDirac}
\end{align}
where $\sigma_{i}$ are the Pauli matrices in the band subspace, $s_{z}$  is the Pauli matrix in the spin subspace, $\sigma_0$ and  $s_{0}$ are unity matrices, $\tau=\pm 1$ is the valley index for $K$ and $K^{\prime} = -K$, $v_{F}$  is the effective Fermi velocity, and $\Delta$ is the bandgap. The last contribution to Eq. (\ref{HDirac}) describes the Zeeman spin-orbit coupling (SOC) with constants $\lambda_{c,v}$. Parameters $v_F$, $\Delta$ and $\lambda_{c,v}$ are obtained by fitting the {\it ab-initio} band structure calculations using  DFT/GW approaches \cite{Zhumagulov2020,Kormnyos2015,Zollner2019}. The gap $\Delta$ depends on the encapsulating materials and needs a correction, e.g. using the scissor operator approach, when one needs quantitatively accurate results\cite{Cho2018}. Finally, the doping-related effects are captured by connecting them with the finite mesh in the $k$ space \cite{Zhumagulov2020,Zhumagulov2020_}. The dipole matrix elements needed to calculate the oscillation strength (OS) of trion states are calculated using obtained trion wave functions.  Further details of the calculation are found in the Supplemental Material.

 \begin{center}
\begin{table}[]
    \centering
    \begin{tabular}{|c|c|c|c|c|c|c|c|}
        \hline
           & $a$ [$\AA$]  & $d$  [$\AA$]  & $\varepsilon_{bulk}$ & $\Delta_0$ [eV]  & $m$  & $\lambda_c$ [meV]  & $\lambda_v$ [meV] \\
         \hline
         \hline
         MoS2&  3.185 & 6.12 & 16.3 & 2.087 & 0.520 & -1.41 &  74.60\\
         \hline
         MoSe2& 3.319 & 6.54 & 17.9 & 1.817 & 0.608 & -10.45 &  93.25\\
          \hline
         WS2 &   3.180  & 6.14 & 14.6 & 2.250 & 0.351 & 15.72  &  213.46\\
          \hline
         WSe2&  3.319 & 6.52 & 16.0 & 1.979 & 0.379 & 19.85  &  233.07\\
         \hline
    \end{tabular}
    \caption{Model parameters table for TMDC ML's. $a$, $m$,  $\lambda_{c,v}$ are taken from Ref. \cite{Zollner2019}, $d$ is from Ref. \cite{Laturia2018} ), and $\Delta_0$ is from Ref. \cite{Zhang2016}  (see model (\ref{HDirac})). }
    \label{tab:parameters}
\end{table}
\end{center}

\begin{figure}
\begin{center}
\includegraphics[width=0.5\textwidth]{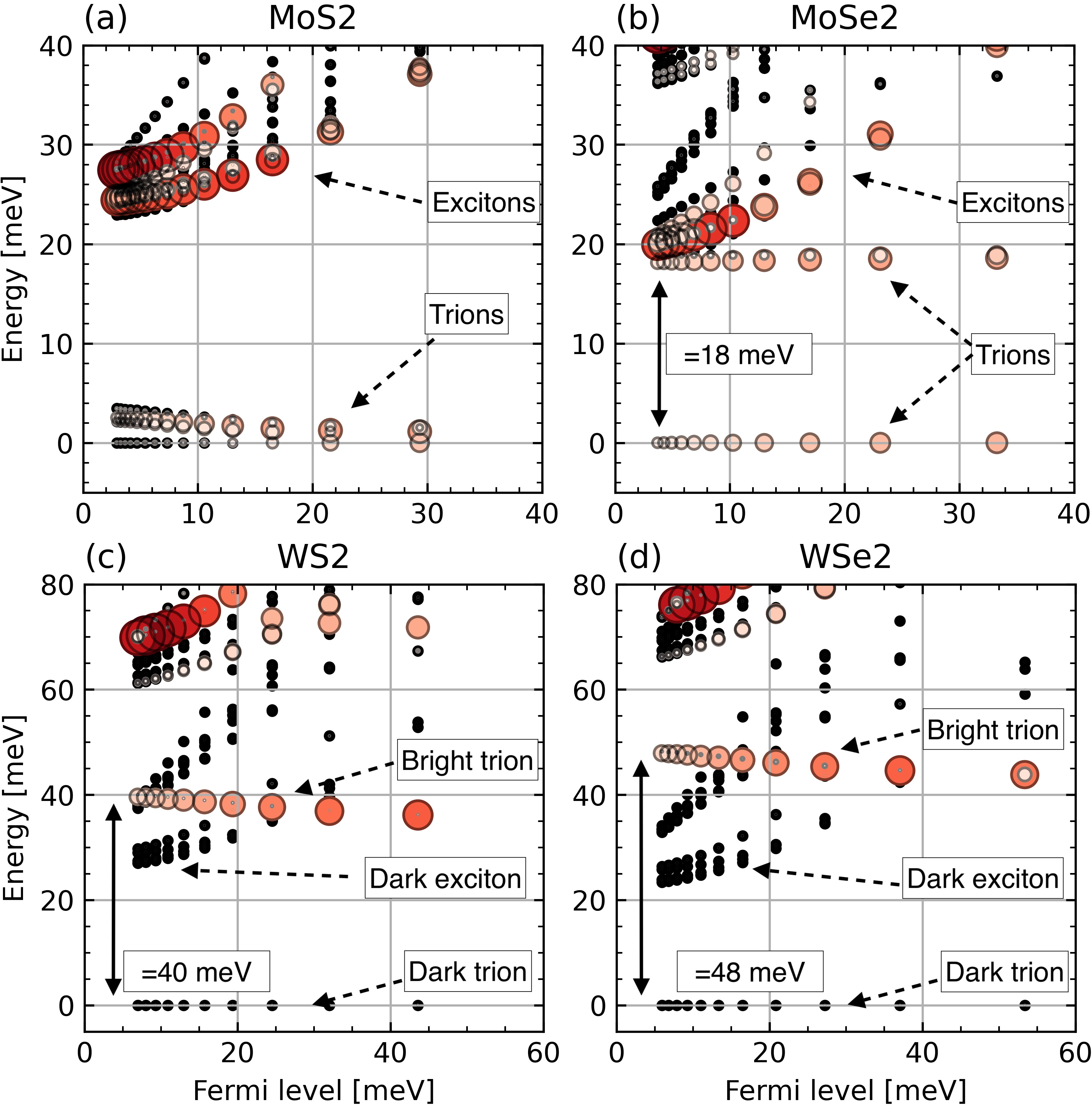}
\end{center}
\caption{Energy diagram of three-particle states calculated freestanding ML's of MoS2 (a), MoSe2 (b),  WS2 (c),  and  WSe2 (d) as a function of the doping Fermi level. Circle positions denote state energies. The colour and width indicate their OS - red for bright states, black  for dark ones. "Trion" and "exciton" denote trion and exciton states (see text).  Only potentially bright states with $\tau s_z=|1/2|$ are shown.}
\label{fig:diag}
\end{figure}

{\it The energy diagram} in Fig. \ref{fig:diag} shows the doping dependence of energies and OS's of three-particle states calculated for freestanding ML's of MoS2, MoSe2, WS2, and  WSe2. In the calculations we use the Dirac model parameters obtained by the {\it ab-initio} calculations and  summarized in Table \ref{tab:parameters}. The color and width of the circles in  Fig. \ref{fig:diag} indicate whether the state is optically active: small dark points denote dark states, and larger circles of red color are bright states, the circle radius gives the state OS.  Note that only potentially bright states with $|\tau s_z| = 1/2$ are shown. The diagram explicitly marks trion states with all electrons tightly bound and excitons with one loose electron.

In all materials, the lowest energy state is a trion (its energy is used as a reference point).  The gap $\Delta E$ between this and the first excited trion state is determined by the SOC. It is large in the W-based materials leading to the large gap,  $\Delta E \simeq 40-50$meV [Figs.  \ref{fig:diag} (c) and (d)]. The SOC is also notable in MoSe2, giving $\Delta E \simeq 20$meV. However, it has a negative value results which means the lowest state in MoSe2 is bright in contrast to the W-based materials. In MoS2, the SOC is small, which makes the energies of the four lowest trion states very close. This near degeneracy opens a possibility to manipulate both the relative energy and the OS of trions by varying system parameters such as doping and dielectric environment. We verified it by observing a large change in the relative energy and OS of the trion states with increasing doping, as shown in Fig. \ref{fig:diag} (a).

\begin{figure}
\begin{center}
\includegraphics[width=0.45\textwidth]{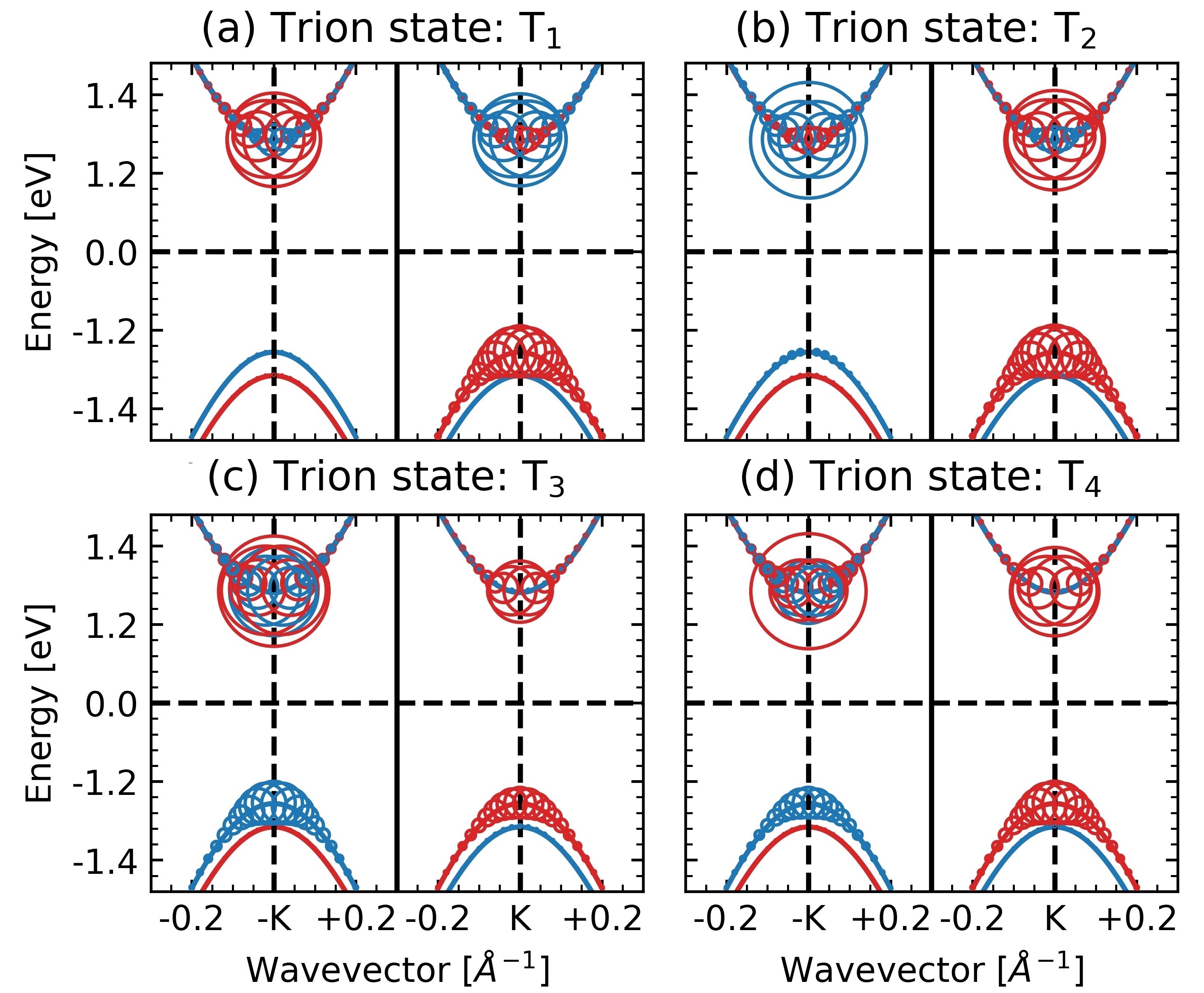}
\end{center}
\caption{Contributions of Dirac single-particle band states in the vicinity of the $K$ and $-K$ points to trions $T_{1–4}$ (panels (a) - (d)). A circle position points to the single-particle state, A circle position points to the single-particle state, and its radius gives the weight to the exciton state, colors mark the spin projection $s_z$. States $T_{1,2}$ in panels (a) and (b) have   $\tau s_z=1/2$, states $T_{3,4}$ in panels (c) and (d) have $\tau s_z=-1/2$. The contributions are shown for $E_F=2.9$ meV.}
\label{fig:trion_wave}
\end{figure}

{\it Internal structure of the trion states} is key to their characteristics.  The two-valley band structure and a strong spin-orbit splitting of the valence and conduction bands predetermine appearance of four types of trion states. Their structure is illustrated in Fig. \ref{fig:trion_wave} that plots the relative contribution of single-particle band states to trions (single-particle density matrix), calculated for the freestanding MoS2 ML at $E_F = 2.9$ meV.  Circles in the figure indicate the contributions with their centers pointing to the contributing single-particle state and the radius giving its weight.  Red and blue color denote the spin $s_z$ of a single-particle state.

Trion states in Fig. \ref{fig:trion_wave} are split into pairs of qualitatively different states, $T_{1,2}$ with $\tau s_z = 1/2$  and $T_{3,4}$ with $\tau s_z = -1/2$. In states $T_{1,2}$ a hole comes from a single valley $-K$, and two electrons of different spin occupy both valleys. The spin of contributing electrons ensures that $T_1$ is dark and $T_2$ is bright (this applies strictly only at vanishing doping). Notice that the transformation $K \leftrightarrow - K$ gives an equivalent trion state. In $T_{3,4}$ states hole states of opposite spin in both valleys contribute.

\begin{figure}
\begin{center}
\includegraphics[width=0.5\textwidth]{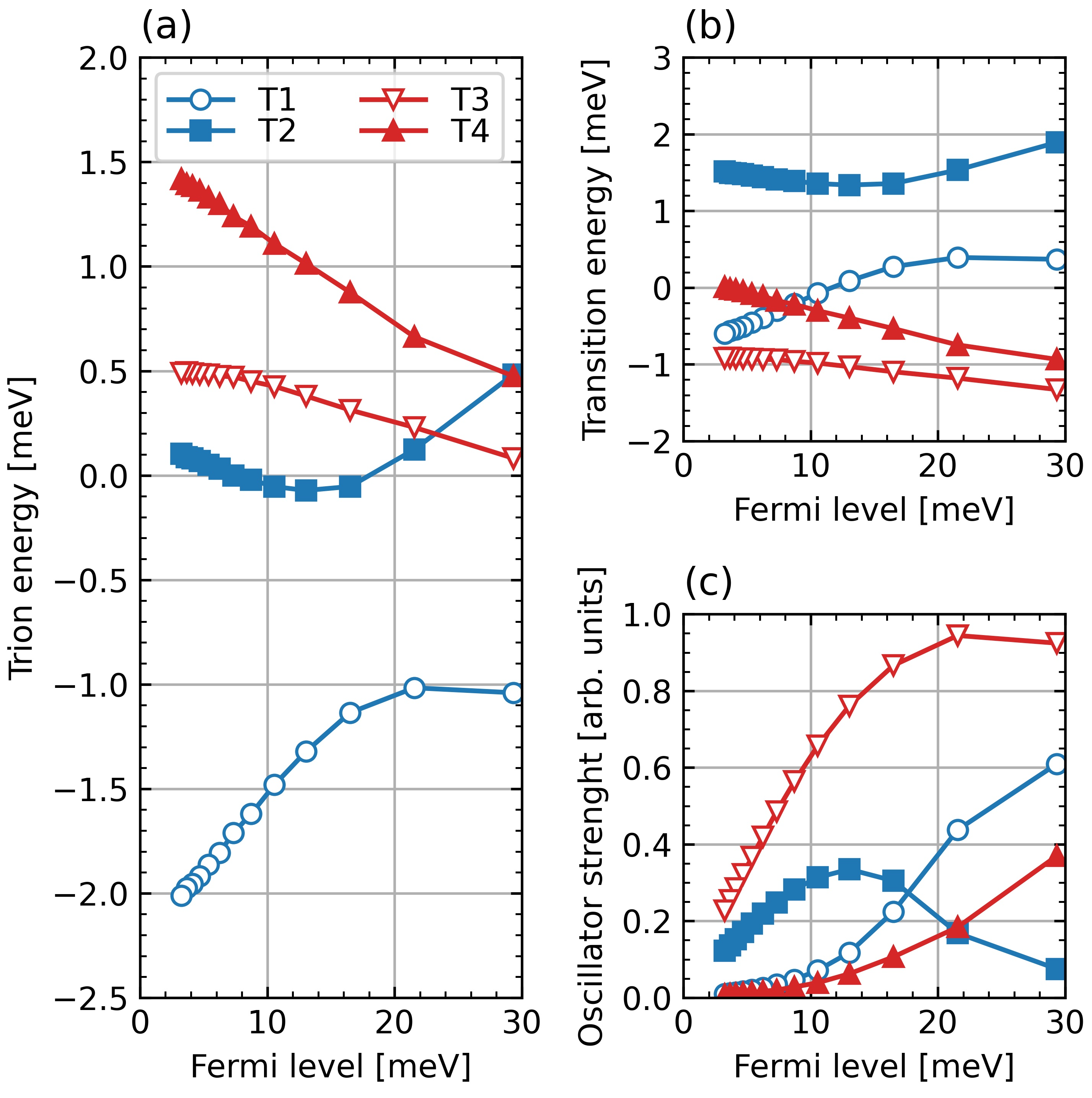}
\end{center}
\caption{The doping dependence of  (a) energies of $T_{1-4}$ trion states of a freestanding MoS2 ML (shifted to have zero mean value),  (b) transition energies of the bright $T_{1-4}$ states (trion energy minus single electron energy), (b) the relative OS of trion states. }
\label{fig:trion_energy}
\end{figure}

\begin{figure}
\begin{center}
\includegraphics[width=0.45\textwidth]{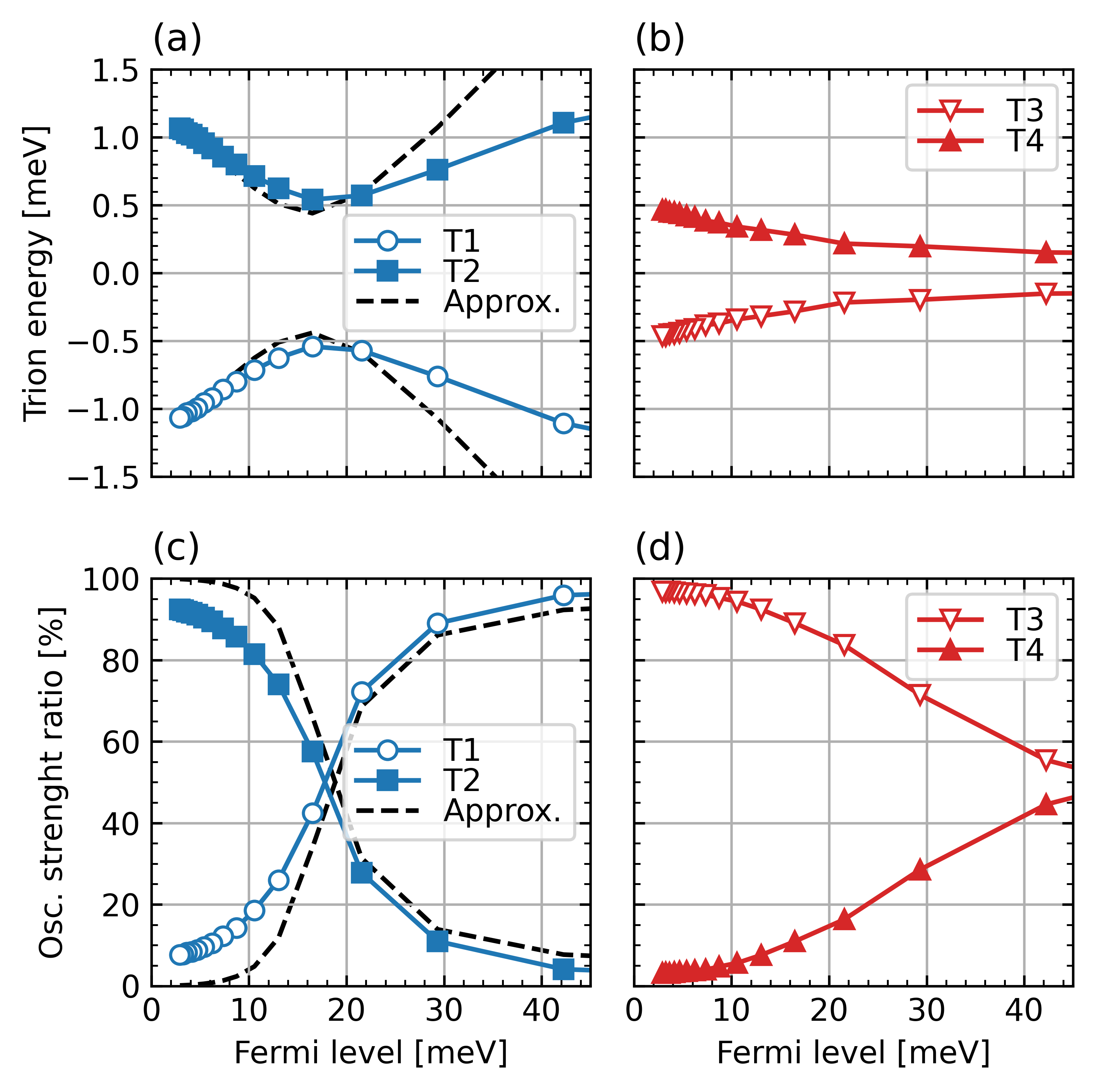}
\end{center}
\caption{Doping dependence of relative energies of trion states  $T_{1,2}$  (a) and $T_{3,4}$ (b), and relative OS of trion states  $T_{1,2}$  (c) and $T_{3,4}$ (d), calculated for a freestanding MoS2 ML. The trion energies are shifted so that their average is zero. The symbols are numerical calculations, dashed black lines are the two-level approximation.}
\label{fig:trion_epsilon_1}
\end{figure}

{\it Doping dependence of the trion states} is illustrated in Fig. \ref{fig:trion_energy} that plots trion energies (a), transition energies (b), and OS's (c). $T_1$ and $T_3$ have the lowest energies in the respective pairs and are dark. Remarkably, the doping dependence of the pair $T_{12}$ is remarkable reveals a clear anti-crossing behavior, both for the energies and OS's, where the brightness of states $T_{1,2}$ is exchanged. The other pair $T_{3,4}$ appears also to have an anti-crossing pattern where the crossing point, however, is shifted to a much higher doping value, $E_F \gtrsim 40$meV.  The anti-crossing pattern is seen clearer in Fig. \ref{fig:trion_epsilon_1} that plots the energy and OS for $T_{1,2}$ (a-b) and $T_{3,4}$ (c-d) with the subtracted average values for clarity. For states $T_{1,2}$ the anti-crossing  takes place at $E_F \simeq 20$meV, whereas for $T_{3,4}$ its position is estimated as $E_F \sim 50$meV.

\begin{figure}
\begin{center}
\includegraphics[width=0.45\textwidth]{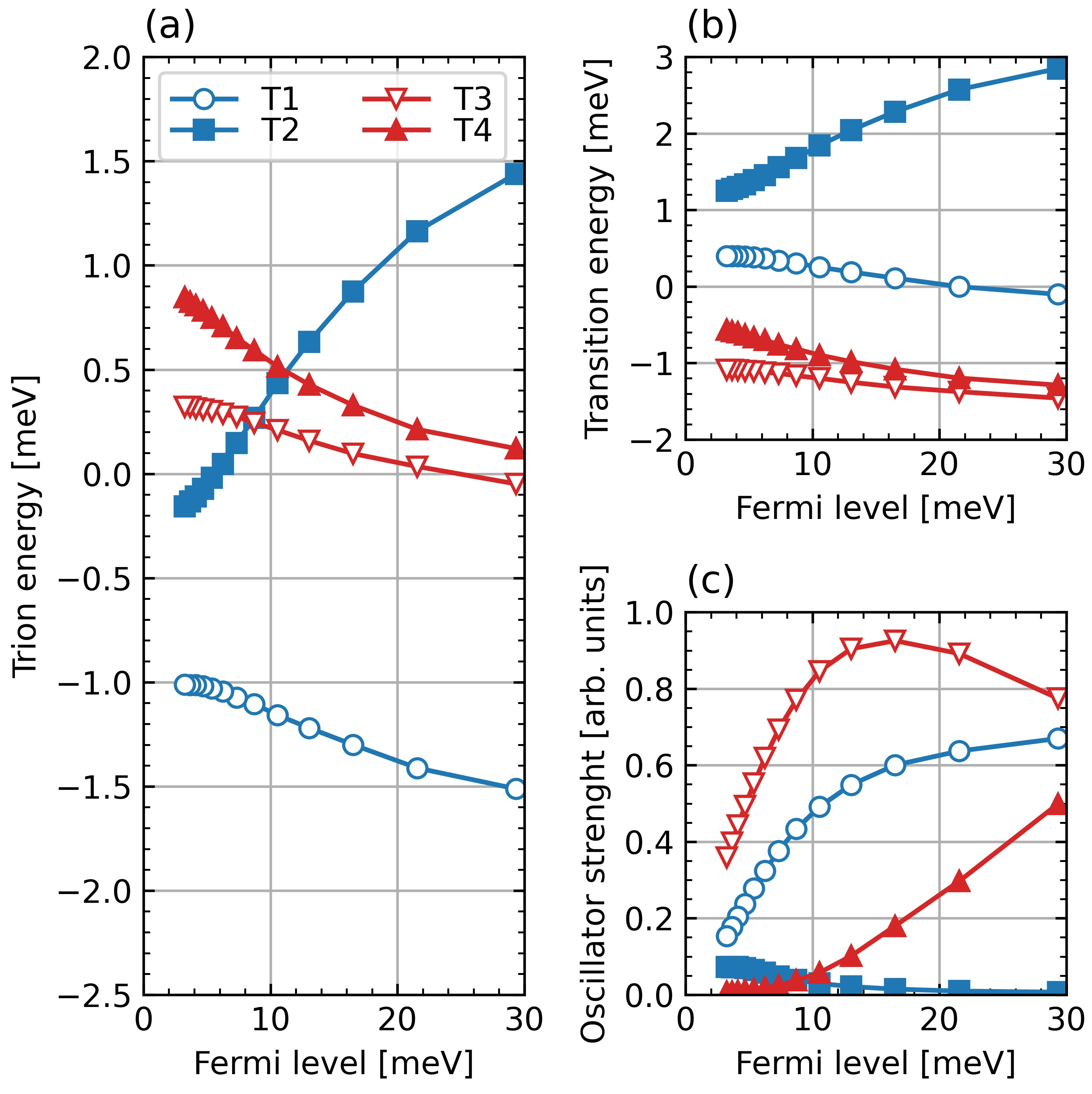}
\end{center}
\caption{Same as in Fig. \ref{fig:trion_epsilon_1} calculated for a MoS2 ML encapsulated in hBN.   }
\label{fig:trion_energy_hBN}
\end{figure}

{\it Dependence of trions on the environment} is presented in Fig.  \ref{fig:trion_energy_hBN} that shows the energy and OS of the trions calculates for a MoS2 ML encapsulated in hBN, which changes the effective dielectric constant $\varepsilon_{env}$ and thus the Coulomb interaction. Values of the energies of $T_{3,4}$ states are slightly shifted, but their relative positions hold being practically the same as for the freestanding ML case [cf Figs.  \ref{fig:trion_energy_hBN} and \ref{fig:trion_energy} (a-b)]. However, the states $T_{1,2}$  are much more sensitive and change qualitatively. The states $T_1$ and T2 interchange their brightness:  $T_1$ is brighter than T2 already at small doping and it becomes even brighter when the doping increases [Fig.  \ref{fig:trion_energy_hBN} (c)]. It appears the anti-crossing still exists, but its position shifts leftwards to negative $E_F$.

Colour-density plots in Fig. \ref{fig:trion_epsilon_1} give further details of the anti-crossing.  Panel (a) gives the difference between the energies of and panel (b) the ratio of the OS for states $T_{1}$ and  $T_{2}$ as functions of the doping Fermi energy and effective dielectric constant. The anti–crossing is given by the line that shows the minimum of the energy difference in Fig. \ref{fig:trion_epsilon_1} (a) as well as by the line where the OS ratio is unity Fig. \ref{fig:trion_epsilon_1} (b).  Both these parameters can effectively control the position relative energy and the brightness of trion states.

{\it   The anti-crossing mechanism} is explained by noting that each trion pair is only weakly coupled to other states and thus can be described by a separate $2\times 2$ model Hamiltonian. For the latter one can use a simple approximation constructed by taking three-particle basis suggested by the actual trion states as shown in Fig. \ref{fig:trion_wave}. Thus, for the $T_{1,2}$ pair we use basis states  $| 1 \rangle = |-K_\uparrow,K_\downarrow,K_\uparrow\rangle$ (dark) and $| 2 \rangle  = |-K_\downarrow,K_\uparrow,K_\uparrow \rangle$ (bright) where contributions with $k \neq \pm K$ is neglected. The matrix Hamiltonian  $\langle i | H| j \rangle$ is calculated using Eq. (\ref{eq:three_particle}). The electron exchange interaction yields the largest contribution to the off-diagonal matrix element $W_{cc}$. The diagonal elements comprise the single- and many-body  components $\epsilon_{i} = \langle i | H| i \rangle =  \epsilon_{i}^{(0)} + \epsilon_{i}^{(1)}$, while their difference $\Delta = \epsilon_{1} - \varepsilon_{2}$ determines the anti-crossing position as the condition $\Delta =0$.  The solution to the Dirac Hamiltonian gives the zero–order values for this quantity $\Delta^{(0)} = 2 \lambda_c$, whereas the electron-hole exchange interaction $V_{cv}$  gives the many-body correction to this quantity, because it is present in $\epsilon_2^{(1)}$ and absent in $\epsilon_1^{(1)}$. This yields  $\Delta = 2 \lambda_c + V_{cv}$.  Eigenstates of this pair Hamiltonian are
\begin{align}
  &  |+\rangle = \cos(\theta/2) |1\rangle + \sin(\theta/2) |2\rangle, \notag \\
  &  |-\rangle = \sin(\theta/2) |1\rangle - \sin(\theta/2) |2\rangle,
\end{align}
where $\theta  ={\rm arccos}(\Delta /D)$, $D=\sqrt{\Delta^2+4 W_{cc}^2}$, and the corresponding eigenvalues (trion energies) are $ 2 \lambda_{\pm}= \epsilon_1 + \epsilon_2 \pm D $. It is then clear that when $\lambda_c$ is large ($\lambda_c \gg W_{cc}$) and positive, dark state $|1 \rangle$ has the lowest energy. This situation takes place in the W-based ML's. In the opposite case of large negative $\lambda_c$, taking place in MoSe2, the lowest energy trion is in the bright $|2 \rangle$ state. The most interesting is the case of  MoS2  when $\lambda_c<0$ and $|2\lambda_c| \sim V_{cv}$. Here we are in the proximity of the crossover point and the trion states can be easily manipulated by changing external parameters. In particular, the doping and effective dielectric constant strongly affect the interaction $V_{cv}$.  Numerical calculations confirm the intuitive expectation that $V_{cv}$ decreases at larger doping resulting in the anticrossing point and the interchange of the OS of the two lowest trion states, $T_{1,2}$.

{\it In summary,} the analysis of trion states from a direct solution of the three-particle Hamiltonian clarifies the spectral features due to the bright and dark trions in TMDC ML's. It turns out that the exchange interaction and spin-orbit coupling determine the energies and oscillator strengths of the trion states. Four existing lowest-energy trion states are split into pairs which, depending on the spin-orbit coupling, reveal a dark-bright anti-crossing pattern. It opens the way to manipulate the position and brightness of the corresponding spectral peaks. In particular, the small spin-orbit coupling in  MoS2 ML's allows one to control the fine structure of trions and the optical spectra by changing electrostatic doping and the dielectric environment. Based on the interplay between the fine-structure splitting and many-body effects, the controlling mechanism is generic and applies to other similar structures. Our results explain existing experimental observations of trion states in 2D multi-valley materials and open new perspectives for their optoelectronic applications.

{\it We acknowledge} support by the Russian Science Foundation under the grant 18-12-00429 used for numerical calculations. Y.\,V.\,Z is grateful to the Deutsche Forschungsgemeinschaft (DFG, German Research Foundation) SPP 2244 (Project-ID 443416183) for the financial support.
V.\,P. acknowledges support from the Vice President for Research and Economic Development (VPRED) and the Center for Computational Research at the University at Buffalo (\url{http://hdl.handle.net/10477/79221})

\begin{figure}
\begin{center}
\includegraphics[width=0.5\textwidth]{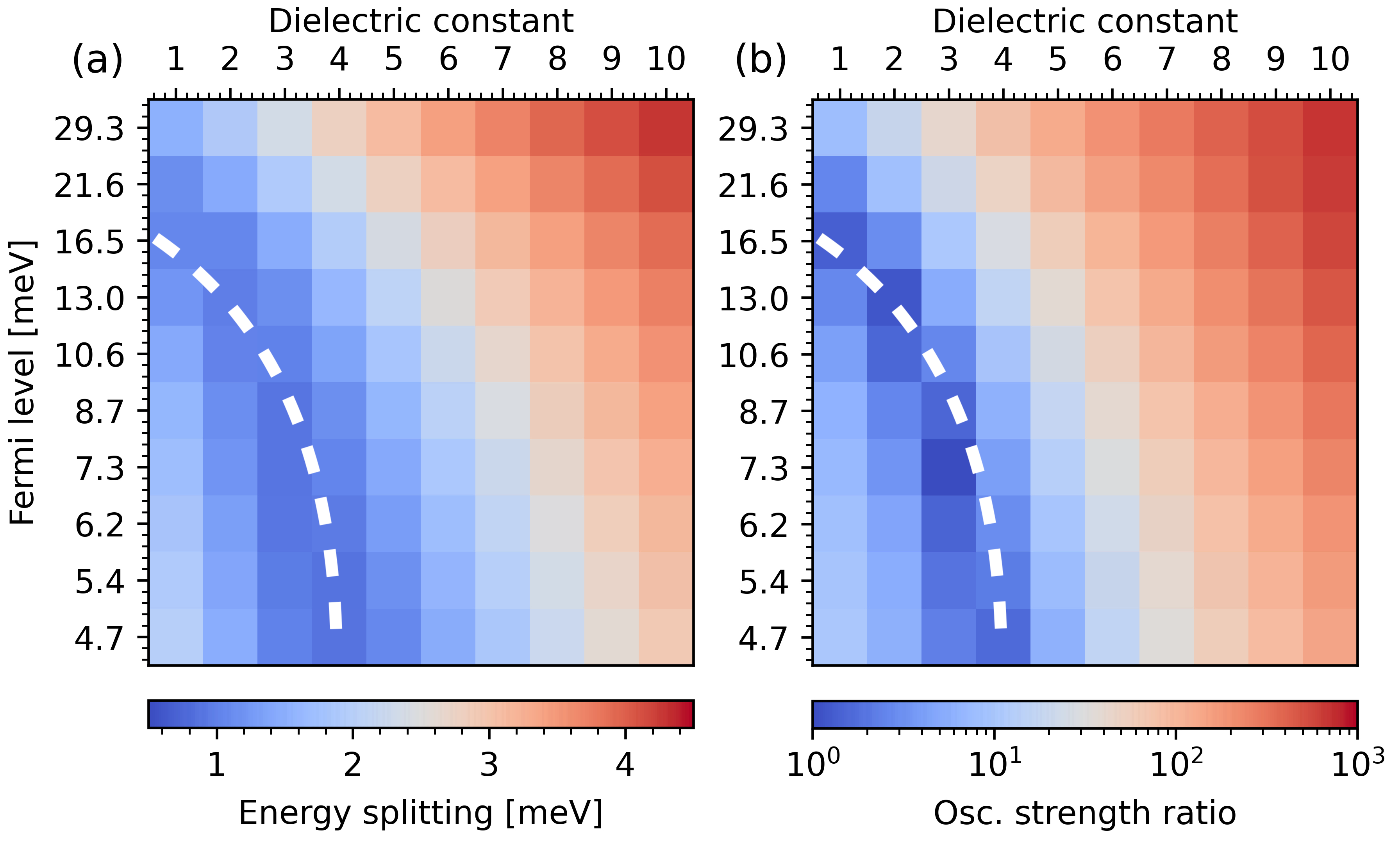}
\end{center}
\caption{(a) Doping/Dielectric dependence of the  energy splitting between $T_{1,2}$ trion states of the MoS2 monolayer . (b) Doping/Dielectric dependence of the oscillation strength ratio of the $T_{1,2}$ trion states of the MoS2 monolayer. The dashed white line corresponds to the anticrossing point on the doping/dielectric diagram.}
\label{fig:colour_density}
\end{figure}

\bibliography{bibliography}

\end{document}


\preprint{APS/123-QED}

\title{Supplemental material for: Electrostatic control of the trion fine structure in transition metal dichalcogenides monolayers}

\author{Yaroslav V. Zhumagulov$^{1,2}$}
\author{Alexei Vagov$^{2}$}
\author{Dmitry R. Gulevich$^{2}$}
\author{Vasili Perebeinos$^{3}$}

\affiliation{$^{1}$University of Regensburg, Regensburg, 93040, Germany}
\affiliation{$^{2}$ITMO University, St. Petersburg 197101, Russia}
\affiliation{$^{3}$Department of Electrical Engineering, University at Buffalo, The State University of New York, Buffalo, New York 14260, USA}

\maketitle

\section{Introduction}

Trion states in a doped monolayer TDMC are calculated by solving an eigenvalue problem for the Hamiltonian for trion states constructed using a basis of single-particles states in the valent and conduction bands. The doping influence is taken into account by employing a method of connecting the Brillouin zone's discretization with the doping-induced Pauli blocking~\cite{Zhumagulov2020}.

\subsection{Single-particle states}

In the calculations, we utilize an observation  that the lowest trion states in TMDC monolayers are created from the single particles states in the vicinity of  points $K$ and $K^\prime = - K$ of the Brillouin zone with the lowest energy~\cite{Drppel2017,Zhumagulov2020}. These states are well described by the massive Dirac Hamiltonian model~\cite{Xiao2012}
\begin{align}
    H_0&=  \hbar v_{F} \big( \tau k_x\sigma_x+k_y\sigma_y \big) + \frac{\Delta}{2} s_0\otimes\sigma_z \notag \\
    &+ \tau s_{z} \otimes \big( \lambda_c \sigma_{+} + \lambda_v \sigma_{-} \big),
\label{HDirac}
\end{align}
where $\sigma_{i,\pm}$ are pseudospin Pauli matrices acting in the band subspace, $s_{z}$  is the spin $z$-component Pauli matrix acting in the spin subspace, and $s_{0}$ is the unity matrix in this subspace, $\tau=\pm 1$ denote valleys $K$ and $K^{\prime}$, respectively, $v_{F}$  is the effective Fermi velocity, and $\Delta$ is the bandgap. The last contribution to Eq. (\ref{HDirac}) describes the Zeeman-type spin-orbit coupling (SOC) with the coupling constants $\lambda_{c,v}$. Despite its relative simplicity, the model captures all relevant phenomena, such as the coupling between the spin and the valley degrees of freedom. Parameters $v_F$, $\Delta$ and $\lambda_{c,v}$ are obtained by comparing the spectrum of the Hamiltonian (\ref{HDirac}) with the band structure of the TMDC monolayer calculated using standard {\it ab initio}  DFT-GW approaches~\cite{Zhumagulov2020}.

We consider that the monolayer is embedded by a dielectric environment, with the dielectric constants $\varepsilon_{1,2}$ for material on both sides of the monolayer.  The environment affects the energy gap $\Delta$, which can be expressed using the analytical formula~\cite{Cho2018}
\begin{align}
    \Delta& =\Delta_{0}+\frac{e^2}{2 \varepsilon d}\left[\frac{L_{2}+L_{1}}{\sqrt{L_{2}L_{1}}}\text{tahn}^{-1}(\sqrt{L_{2}L_{1}}) \right. \notag \\ & \left. -\text{ln}(1-L_{2}L_{1}) \right], \quad L_{i} =  \frac{\varepsilon - \varepsilon_{i}}{\varepsilon + \varepsilon_{i}},
\end{align}
where $\varepsilon$ is the dielectric constant of the bulk TMDC, $d$ is the monolayer thickness and $\Delta_0$ is the bare values of the gap. It is known  that the dielectric environment does not change the effective mass of the single-particle states~\cite{Waldecker2019}. To reflect this fact, the effective Fermi velocity is chosen as  $v_{F}= \sqrt{\Delta/2m}$.

\subsection{Trion states}

Quantum states $T$ of negatively charged trions (two electrons and one hole) are obtained by solving the three-body eigenvalue problem derived by spanning the full many-body Hamiltonian onto the space of trion states constructed as a linear superposition
\begin{align}
  \left| T \right\rangle = \sum_{c_1,c_2,v} A_{c_1 c_2 v}^{T}  \left| c_1 c_2 v \right\rangle, \,  \left| c_1 c_2 v \right\rangle =  a_{c_1}^\dagger  a_{c_2}^\dagger a_{v}^\dagger \left| 0 \right\rangle,
  \label{eq:expansion}
\end{align}
where $c_{1,2}$ denote electron states in the conduction band, $v$ are hole states in the valence band, and the double counting is avoided by imposing the restriction $c_1<c_2$. The corresponding three-particle wavefunction is constructed from the single-particle functions $\phi_{c,v}(x)$ as
\begin{align}
    \Psi^{T}(&x_1,x_2,x_3)=\frac{1}{\sqrt{2}} \sum_{c_1,c_2,v} A_{c_1 c_2 v}^{T} \phi_{v}^{*}(x_3) \notag \\
    &\times \big[ \phi_{c_1}(x_1)\phi_{c_2}(x_2)-\phi_{c_2}(x_1)\phi_{c_1}(x_2) \big].
\label{eq:wave_function}
\end{align}
Coefficients $A_{c_1 c_2 v}^{T}$ in the expansion are found by solving the matrix eigenvalue problem
\begin{align}
\sum_{c_1^\prime c_2^\prime v^\prime} {  H}_{c_1c_2v}^{c_1^\prime c_2^\prime v^\prime} A_{c_1^\prime c_2^\prime v^\prime}^{T} = E_t A_{c_1c_2 v}^{T}.
\label{eq:hamiltonian}
\end{align}
The Hamiltonian matrix comprises three contributions
\begin{align}
& {  H}  = {  H}_{0} + {  H}_{cc}  + {  H}_{cv},  \\
& {  H}_{0}= (\varepsilon_{c_1} + \varepsilon_{c_2} -\varepsilon_{v}) \delta_{c_1c_1^\prime} \delta_{c_2c_2^\prime}  \delta_{v v^\prime}, \nonumber \\
& {  H}_{cc} = (W_{c_1c_2}^{c_1'c_2'}-W_{c_1c_2}^{c_2'c_1'})\delta_{vv'}, \nonumber \\
& {  H}_{cv}= -(W_{v'c_1}^{vc_1'}-V_{v'c_1}^{c_1'v})\delta_{c_2c_2'}-(W_{v'c_2}^{vc_2'}-V_{v'c_2}^{c_2'v})\delta_{c_1c_1'} \nonumber \\
& \quad +(W_{v'c_1}^{vc_2'}-V_{v'c_1}^{c_2'v})\delta_{c_2c_1'}+(W_{v'c_2}^{vc_1'}-V_{v'c_2}^{c_1'v})\delta_{c_1c_2'}, \nonumber
\end{align}
where $\varepsilon_{c,v}$ denote the energy of single-particle particle (hole) states, $W$ and $V$ are the screened and bare Coulomb matrix elements, respectively. This approach is a direct extension of the Tamm-Dancoff approximation for two-particle excitons onto three-particle trions. Matrix elements for the bare Coulomb interaction are given by
\begin{equation}
\label{eq:Coulomb}
    V^{ab}_{cd} = V(\textbf{k}_a -\textbf{k}_c) \langle u_c | u_a \rangle \langle u_d | u_b \rangle,
\end{equation}
where $V (q) = 2\pi e^2/ q$ is the Fourier component of the bare Coulomb potential and $\langle u_c| u_a \rangle$ is the overlap of the single-particle Bloch states $c$ and $a$. The screened potential is also given by Eq. (\ref{eq:Coulomb}), however, instead of the bare Coulomb potential $V(q)$ one substitutes the standard Rytova-Keldysh  expression~\cite{rytova1967the8248,keldysh,Cudazzo2011}
\begin{align}
\label{eq:Coulomb_screened}
   W(q)=
   \frac{2\pi}{q \varepsilon_{env}} \begin{cases}
    (1+r_0 q)^{-1} & q-\text{intravalley} \\
   1 &  q-\text{interavalley} \\
    \end{cases} ,
\end{align}
where the average dielectric constant is $\varepsilon_{env} = (\varepsilon_{2}+\varepsilon_{1})/2$ and  the screening length is given by $r_0=\varepsilon d/2$~\cite{Berkelbach2013,Cho2018}. Notice that the screening depends on device geometry. In particular, it is modified when the vacuum spacing appears between the monolayer and the dielectric environment~\cite{Florian2018}. However, in this work, this effect is neglected.

Symmetries of the three-particle Hamiltonian give rise to several conserved quantities, which helps to reduce the numerical efforts. These include the total trion momentum ${\bf k} = {\bf k}_{c_1} + {\bf k}_{c_2} - {\bf k}_v$ and $z$ axis spin projection $ s_{z} = s_{z}^{c_1} + s_{z}^{c_2} - s_{z}^v$. Furthermore, the analysis is restricted to trions with a total spin $1/2$, as only such states are optically active states allowing for the radiative recombination of electron-hole pairs. For example, we neglect the states with the spin $3/2$ (transitions between spin $1/2$ and $3/2$ states can be facilitated by magnetic impurities, neglected in this work).

The trion states are classified by the $z$ component of the electron spin, $s_{z}^{c_1} + s_{z}^{c_2}=1$ and $0$~\cite{Yu2014,Mayers2015,Courtade2017,Torche2019,Tempelaar2019}, and distinguish between the intra- and inter-valley trion states by the quantity  $\tau_{c_1} + \tau_{c_2}=2$ and $0$, respectively~\cite{Torche2019,Tempelaar2019}. Trion states can also be classified by the product $s_{z}  \tau$ of the total spin projection $s_{z}$ and the total valley index $\tau = \tau_{c_1} + \tau_{c_2} - \tau_v$. A necessary (but not sufficient) condition for a state being bright is $s_{z} \tau =   \pm 1/2$.

Finally, the influence of electron doping is taken into account by using the approach described in Ref.~\cite{Zhumagulov2020}, where the Pauli blocking of the occupied electron states is modeled by choosing the discretized mesh with the interval $\delta K$ in the Brillouin zone. The doping density  $n=g_v g_s/(\Omega_0 N^2)$ is related to the number of mesh points $N\times N$ and the area of the primitive cell $\Omega_0$ ($g_s$ and $g_v$  are the spin and valley degeneracies, respectively). The corresponding Fermi energy of the doping electrons is given by $E_F= \delta k^2/2m$ where the discretization interval for the hexagonal lattice of MoS2 is $\delta k = 4\pi/(\sqrt{3}aN)$.

\subsection{Trion $s_z\tau=\pm3/2$ states}
\begin{figure}
\begin{center}
\includegraphics[width=0.75\textwidth]{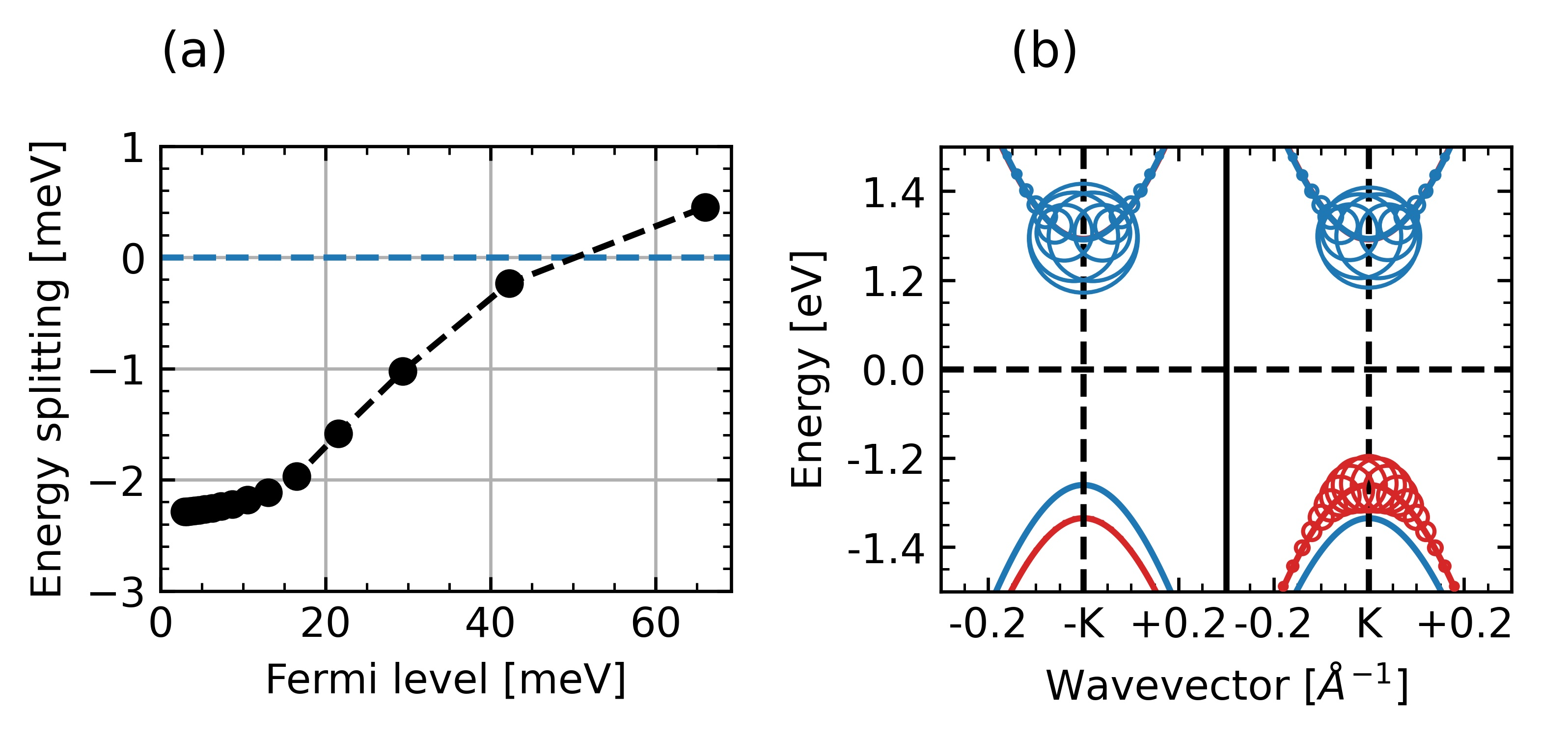}
\end{center}
\caption{(a) Trion energy splitting between the lowest $s_z\tau=\pm3/2$ state and the lowest $s_z\tau=\pm1/2$ states of the freestanding of MoS2 monolayer. For a wide range of doping, fully dark $s_z\tau=\pm3/2$ state will be the lowest trion state of the freestanding of MoS2 monolayer. (b) Contributions of Dirac single-particle band states in the vicinity of the $K$ and $-K$ points to the lowest $s_z\tau=\pm3/2$ trion state.}
\label{fig:s1}
\end{figure}

The trion states $s_z\tau=\pm3/2$ are fully dark and scattering between  $s_z\tau=\pm3/2$ and  $s_z\tau=\pm1/2$ requires a nonzero spin-boson. However, in the presence of magnetic impurities, it possible to obtain $s_z\tau=\pm3/2$ trion state in the freestanding of MoS2 monolayer as the lowest trion states, with significant changes in  photoluminescence temperature dependence. Energy splitting between the lowest $s_z\tau=\pm3/2$ state and the lowest $s_z\tau=\pm1/2$ state will be about 2.3 meV in the zero doping level limit, as shown in Fig.~\ref{fig:s1}a. This effect is due to the absence of repulsive exchange interaction between the hole and electrons, see Fig.~\ref{fig:s1}b, and will also be observed only in the MoS2 monolayer due to low conduction band splitting, which is comparable with the amplitude of the exchange interaction.

\bibliography{bibliography}